\documentclass[journal]{IEEEtran}
\usepackage{cite}
\usepackage{amsmath,amssymb,amsfonts}
\usepackage{textcomp}
\usepackage{threeparttable}

\usepackage{amsthm}
\usepackage{setspace}

\usepackage{algorithm}
\usepackage{algpseudocode}

\usepackage{epstopdf}
\usepackage{epsfig}
\usepackage{amsmath, amsfonts}
\usepackage{latexsym, bm}

\usepackage{mathrsfs}
\usepackage{amsbsy}

\usepackage{graphicx}
\usepackage[caption=false]{subfig}

\usepackage{algorithm}
\usepackage{algpseudocode}

\usepackage{color}

\usepackage{cite}

\begin{document}

\title{A Scalable 256-Elements E-Band Phased-Array Transceiver for  Broadband Communications}


\author{Xu Li, Wenyao Zhai, Morris Repeta, Hua Cai, Tyler Ross, Kimia Ansari, Sam Tiller, Hari Krishna Pothula,  Dong Liang, Fan Yang, Yibo Lyu, Songlin Shuai, Guangjian Wang,  Wen Tong
\thanks{X. Li, H. Cai, D. Liang, F. Yang, S. Shuai, G. Wang are with the Wireless Technology Laboratory, Central Research Institute, Huawei Technologies Co., Ltd., China.  (e-mail: \{lixu11, bruce.caihua, richard.liangdong, yangfan184, lvyibo, shuaisonglin, wangguangjian\}@huawei.com). }
\thanks{W. Zhai, M. Repeta, T. Ross, K. Ansari, S. Tiller, H. K. Pothula,  W. Tong are with the Huawei Technologies Canada Research Center, Ottawa, Ontario, Canada. (e-mail: \{wenyao.zhai, morris.repeta, tyler.ross, kimia.ansari, sam.tiller, hari.krishna.pothula, tongwen\}@huawei.com).}
\thanks{Corresponding author: M. Repeta (E-mail:morris.repeta@huawei.com), W. Zhai (E-mail:wenyao.zhai@huawei.com).}
\thanks{The first four authors contribute equally to this paper.}}


\maketitle
\section{Abstract}
For E-band wireless communications, a high gain steerable antenna with sub-arrays is desired to reduce the implementation complexity. This paper presents an E-band communication link with 256-elements antennas  based on 8-elements sub-arrays and four beam-forming chips  in silicon germanium (SiGe) bipolar complementary metal-oxide-semiconductor (BiCMOS), which is packaged on a 19-layer low temperature co-fired ceramic (LTCC) substrate. After the design and manufacture of the 256-elements antenna, a fast near-field calibration method is proposed for calibration, where a single near-field measurement is required. Then near-field to far-field (NFFF) transform and far-field to near-field (FFNF) transform are used for the bore-sight calibration. The comparison with high frequency structure simulator (HFSS) is utilized for the non-bore-sight calibration. Verified on the 256-elements antenna, the beam-forming performance measured in the chamber is in good agreement with the simulations. The communication in the office environment is also realized using a fifth generation (5G) new radio (NR) system, whose bandwidth is 400 megahertz (MHz) and waveform format is orthogonal frequency division multiplexing (OFDM) with 120 kilohertz (kHz) sub-carrier spacing.

\begin{IEEEkeywords}
 Calibration, E-band,  LTCC, SiGe BiCMOS, Sub-array.
\end{IEEEkeywords}

\section{Introduction}
Cellular data rate demand is growing at an exponential rate, to tens of Gigabits per second (Gbps) in the fifth generation (5G) new radio (NR) communication \cite{wenyao_1} and even Terabits per second (Tbps) in the future sixth generation (6G) communication \cite{6G}. A promising method to increase the data rate is to increase the spectrum bandwidth, where E-band with 10 gigahertz (GHz) spectrum available is a candidate \cite{wenyao_2}. To overcome the high propagation loss at E-band,  a large-scale high gain and steerable phased array antenna is required to improve the link performance \cite{wenyao_3}. The large-scale phased array antenna, requires complicated control circuits which is bulky and expensive. To reduce the hardware complexity, the sub-array architecture is developed.

\subsection{Contributions}
The methodology of a E-band limited field of view (LFOV) phased array using rectangular sub-arrays with steering range of $\pm 15^\text{o}$ in  azimuth plane is designed in our previous works in \cite{our_00, our_000} in 2016 and 2017.  This paper presents a 256-elements antenna, based on 8-elements sub-arrays and beam-forming chips in silicon germanium (SiGe) bipolar complementary metal-oxide-semiconductor (BiCMOS) packaged on a 19-layer low temperature co-fired ceramic (LTCC) substrate, which is briefly introduced in \cite{our_1}. The 256-elements antenna has 32 sub-arrays, where each has 8 elements with four possible configurations: $1\times8$, $8\times1$, $2\times4$ and $4\times2$. Meanwhile, there are four Active Antenna (AA) and one Up/Down Converter (UDC) ASICs. The AA ASIC receivers/transmits E-band signal to/from the sub-array, which has eight transmiter/receiver (TRx) chains. The UDC ASIC transmits/receives intermediate frequency (IF) signal from the baseband module, and up-converts/down-converts E-band signal to/from AA chips, which can feed eight AAs.

Then the amplitude and phase vary among different chains due to non-ideal circuits, which degrades the beam-forming performance.  A fast near-field calibration method is proposed, which makes bore-sight calibration and non-bore-sight calibration sequentially. For the bore-sight calibration, similar to \cite{papers_1, papers_2}, a single near-field measurement, near-field to far-field (NFFF) transform and far-field to near-field (FFNF) transform are used. To improve the accuracy, the influence to each sub-array from adjacent sub-arrays are removed. For the non-bore-sight calibration, the comparison with high frequency structure simulator (HFSS) is used along with the bore-sight calibration result. Validation on our antenna measured in the chamber, it restores the array beam-forming performance in good agreement with the simulations, and reduces the calibration time from $40 \times 32$ minutes to $40$ minutes.

Moreover, a test demo in the office environment is  realized using a  5G NR system to evaluate the performance of the 256-elements antenna, where the bandwidth is 400 megahertz (MHz) and waveform format is orthogonal frequency division multiplexing (OFDM) with 120 kilohertz (kHz) sub-carrier spacing

\begin{table*}[]
\centering
\caption{The mm-Wave phased array with implementation.}
\label{tb00}
\renewcommand{\arraystretch}{1.0}
\begin{tabular}{p{0.1\textwidth}|p{0.1\textwidth}|p{0.1\textwidth}|p{0.13\textwidth}|p{0.13\textwidth}|p{0.1\textwidth}|p{0.08\textwidth}}
\hline
Paper & Number of element & Number of sub-arrays & Frequency  & EIRP/Gain & Worst SLL  & Aperture efficiency  \\
\hline
\cite{wenyao_new_16} & 64 & 64 & 28-32 GHz  & 20 dBm EIRP &  $-20$ dBc  & -  \\
\hline
\cite{wenyao_new_15} & 100 & 64 & 75-110 GHz  & 24 dBm EIRP &   -  & -  \\
\hline
\cite{wenyao_new_14} & 144 & 72 & 57-66 GHz  & 51 dBm EIRP &  $-10$ dBc  & -  \\
\hline
\cite{wenyao_new_13} & 392 & 196 & 8-12 GHz  & 22.5 dBi Gain &  $-12$ dBc  & $53.4\%$  \\
\hline
\cite{wenyao_new_12} & 256 & 256 & 58-63 GHz  & 26 dBi Gain &  $ -12$ dBc  & $ 50\%$  \\
\hline
This work & 256 & 32 & 71-76 GHz  & 51 dBm EIRP &  $-12$ dBc  & $ 50\%$  \\
 &  &  &   & 26 dBi Gain &   &   \\
\hline
\end{tabular}
\end{table*}

\begin{table*}[]
\centering
\caption{The near-field calibration method with implementation.}
\label{tb0}
\renewcommand{\arraystretch}{1.0}
\begin{tabular}{p{0.3\textwidth}|p{0.263\textwidth}|p{0.26\textwidth}}
\hline
Near field measurement once  & Without or identical sub-arrays & Randomly tiled sub-arrays\\
\end{tabular}
\begin{threeparttable}
\begin{tabular}{p{0.14\textwidth}|p{0.14\textwidth}|p{0.12\textwidth}|p{0.12\textwidth}|p{0.12\textwidth}|p{0.12\textwidth}}
\hline
Paper & Frequency & Bore-sight & Non-bore-sight  & Bore-sight & Non-bore-sight  \\
\hline
\cite{nearfield_1} & 100 kHz & yes\tnote{} & yes  & - & -  \\
\hline
\cite{nearfield_2} & 1.5-2.5 GHz & yes & yes  & - & -  \\
\hline
\cite{nearfield_3} & 5.34-5.46 GHz & yes & yes  & - & -  \\
\hline
\cite{nearfield_4} & 8-12 GHz & yes & yes  & - & -  \\
\hline
\cite{papers_2} & 38.5 GHz & yes & yes  & yes & no  \\
\hline
This work & 72.2 GHz & yes & yes  & yes & yes  \\
\hline
\end{tabular}
 \begin{tablenotes}
        \footnotesize
        \item[] yes/no: It can/can't support the calibration for specified steering directions and sub-carry type.
        \item[] -: Sub-array structure is not required at low frequency.
      \end{tablenotes}
    \end{threeparttable}
\end{table*}


\subsection{Related Works}
\subsubsection{Sub-array} The steering range of a sub-array based phased array is critical, which can be improved  by breaking  the periodicity of sub-array tiling of an array aperture. There are numerous methods, where the first category is aperiodic sub-array tiling with irregular shape sub-arrays, such as polyomino \cite{wenyao_new_1}, diamond \cite{wenyao_new_2}, random size \cite{wenyao_new_3} and poly-hexagon \cite{wenyao_new_4}.  Additional control circuits are not needed to drive each sub-array, and it is possible to use big size sub-arrays. But the feeding network of the phased array is difficult to be implemented. The second category consists of manipulation of regular shape sub-arrays, such as randomize tiling of sub-array of two or four \cite{wenyao_new_5}, sub-array rotation \cite{wenyao_new_7} and overlap sub-arrays \cite{wenyao_new_8}. The design of the sub-array tiling and the implementation of feeding network are easier than that with irregular shape sub-arrays. But additional circuits are needed, such as power combiner/splitter or cross-over, wherever there is an overlapped element between sub-arrays. The last category is sub-array based adaptive processes, where the element/sub-array radiation patterns \cite{wenyao_new_9}, complex excitations of sub-arrays \cite{wenyao_new_10}, or additional phase adjustment of a random element within a sub-array \cite{wenyao_new_11} are adaptively optimized over array steering range. However, all these works are performed analytically without implementation to proof the concept. The proposed phased array is compared against other start-of-arts mm-Wave phased array modules  \cite{wenyao_new_16,wenyao_new_15,wenyao_new_14,wenyao_new_13,wenyao_new_12} as shown in Table \ref{tb00}. The proposed phased array using rectangular sub-arrays has the highest element-control ratio, so that the circuit implementation is simpler with more module area available for heat dissipation, and  decent array gain, SLL and overall aperture efficiency are achieved.
\subsubsection{Calibration}
The calibration method can be implemented with circuit in the radio frequency (RF) chains \cite{circuitry_1, circuitry_2}. Based on the feedback signal from the calibration circuitry, the variations can be corrected. However, the calibration circuitry increases the design and manufacturing complexity. Alternative calibration methods use far-field and near-field measurements, whose calibration probe is placed in the far-field \cite{farfield_1, farfield_2} and near field areas  \cite{papers_2, nearfield_1, nearfield_2,nearfield_3,nearfield_4}. Another option is mutual coupling measurement, where part of the array aperture is used as the calibration probe \cite{mutual_coupling_1, mutual_coupling_2}. Commonly, the probe collects signal from one antenna while turning off others.  However at E-band with wavelengths around $4$ mm, the calibration probe needs to be positioned with high accuracy \cite{position_1}. A $1$ mm position error can lead to a $90^o$ calibration phase error. Moreover, the radiation pattern of randomly tiled sub-arrays are not unique, so that the far-field measurement and mutual coupling may not work. The proposed near-field calibration method is compared against other ones \cite{nearfield_1, nearfield_2,nearfield_3,nearfield_4} with implementation as shown in Table \ref{tb0}. The proposed calibration method can tackle the non-bore-sight calibration with randomly tiled sub-arrays.


The remainder of this paper is organized as follows. Section \ref{chip_design} describes chip design of the 256-elements antenna. Section \ref{ltcc} introduces  LTCC of the 256-elements antenna. Section \ref{cali_1} presents the calibration for bore-sight, followed by the calibration for non-bore-sight in Section \ref{cali_2}. Section \ref{results} shows the results with our manufactured 256-elements antenna, the test system, the beam-forming calibration performance in the chamber, and the over the air (OTA) communication performance in the office environment. Section \ref{conclusion} concludes the paper.

\begin{figure}[]
\centering
\includegraphics[width=0.5\textwidth]{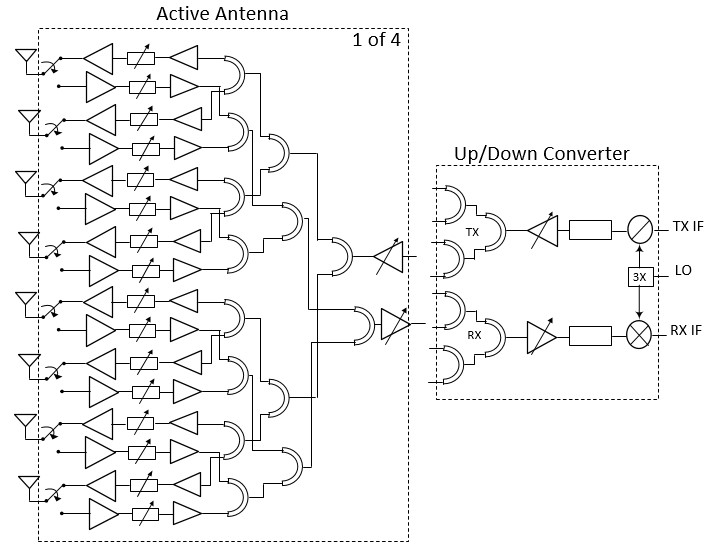}
\caption{E-band phased-array transceiver architecture.}
\label{fig:ims-fig1}
\end{figure}

\section{Beam-Former Chip Design} \label{chip_design}
As illustrated in Fig. \ref{fig:ims-fig1}, the system includes two kinds of chips;  an AA and an UDC  ASICs .



\subsection{AA ASIC}

The AA ASIC receivers/transmits E-band signal to/from the sub-array, which has eight TRx chains. It contains power combiners/splitters, a low-noise amplifiers (LNA), two 5-bits amplitude adjustable power amplifiers (PAs) and two 6-bits phase shifters (PSs). The PAs and PSs are controlled using an SPI interface for variable gain and phase, to achieve the beam-forming and shaping.  Since our system is intended to support a  time division multiplexing (TDD) system, RF switches are connected to antenna so that the antennas are shared between Rx and Tx.
More details regarding the design of mm-wave building blocks can be found in \cite{ims2,ims3,ims4}.  The size of the AA ASIC is measured as $4.4 \times 6.8 \text{ mm}^2$, where the power consumption of the eight Tx chains is 1 W, and that of the eight Rx chains is 0.5 W.




\subsection{UDC ASIC}
The UDC ASIC transmits/receives IF signal from the baseband module, and up-converts/down-converts E-band signal to/from AA chips, which can feed up to eight AAs \footnote{The last stage of combiner/splitter is  not drawn for brevity, and four AAs are used in our phased array.}. It contains power combiners/splitters, a frequency tripler, an up-conversion mixer, a down-conversion mixer and two 5-bits variable gain amplifiers  (VGAs) controlled using an SPI interface. The local oscillator (LO) connects to the tripler, and drives the mixer. In the transmitter chain (Tx), the IF signal is up-converted to the E-band signal. The E-band signal is amplified, and split to feed four AAs. In the receiver chain (Rx), the E-band signal from four AAs are combined, amplified, and down-converted to the IF signal. The size of the UDC ASIC is measured as $2.8 \times 3.7 \text{ mm}^2$, where the power consumption of the TRx chains is 200 mW.

\section{LTCC Antenna Design} \label{ltcc}
A 256-elements antenna was designed using Kyocera GL-773 process. The substrate has 19 layers, the total thickness including BGA balls is 2.4 mm and the dimensions are $44 \times 44 \text{mm}^2$. Five layers are used to define the cavity and four for the antennas.  The rest is dedicated to power, IF signal and antenna array feeding network routing as well as matching network at interfaces between ASICs to the LTCC package and between LTCC BGA and mother board.

\begin{figure}[]
\centering
\subfloat[]{
\label{f1}\includegraphics[width=0.24\textwidth]{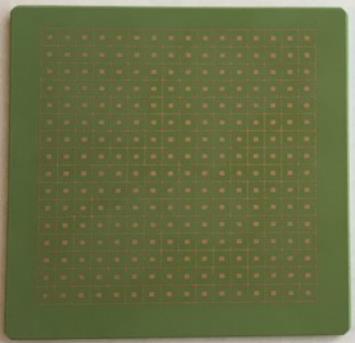}}
\subfloat[]{
\label{f2}\includegraphics[width=0.24\textwidth]{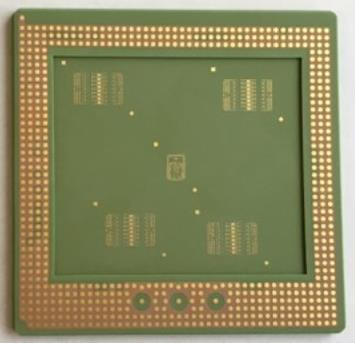}}
\caption{256-elements LTCC substrate, (a) top side showing antennas and (b) bottom side showing BGA-balls and cavity housing four AAs and one UDC.}
\label{fig:ims-fig2}
\end{figure}

%

\begin{figure}[]
\centering
\includegraphics[width=0.5\textwidth]{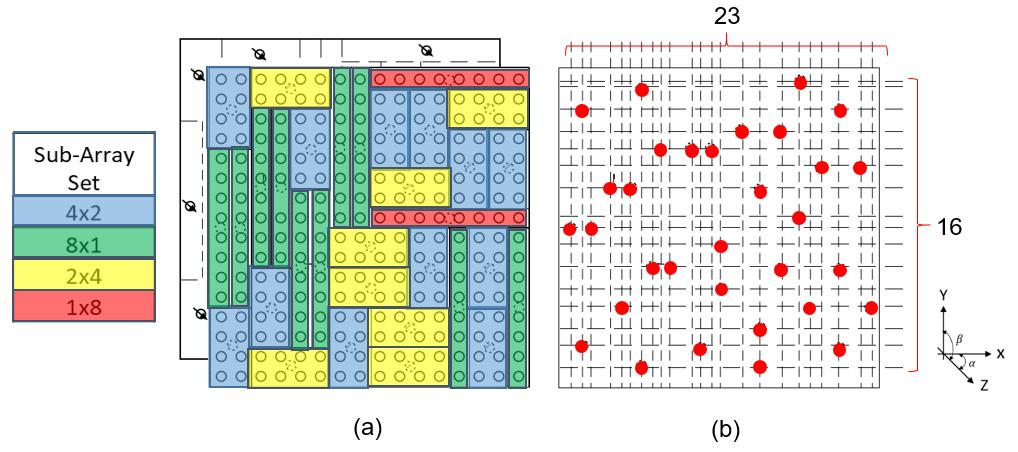}
\caption{Sub-array tiling: (a) randomly tiled 256-elements phased-array with 4 different 8-elements sub-array configurations, and (b) phase shifters granularity of the phased-array aperture in x-y plane.}
\label{fig:ims-fig3}
\end{figure}

Due to constraints such as cost, power and system complexity, it is not feasible to consider having a dedicated transceiver for each antenna element. Sub-arrays are inevitable, so in this work sub-arrays of eight elements were used.  Since each AA comprises eight TRx, four AAs are used as shown in Fig. \ref{fig:ims-fig2}b.  This addresses the constraints mentioned above but high sub-array spacing and periodicity results in grating lobe (GLL) when steering.  This is especially true if periodicity of the sub-array is well determined.

Randomly tiled rectangular sub-arrays are used as discussed in \cite{our_000}. Four configurations are possible: $1\times8$, $8\times1$, $2\times4$ and $4\times2$. Sufficient randomness and granularity can be achieved to get limited field of view (LFOV) performance in both Azimuth and Elevation planes scan range.

Fig. \ref{fig:ims-fig3}a shows a 256-elements phased-array, white dots represent the location of the array elements spaced by $\lambda_0/2$ at 73 GHz. Grey dots indicate the location of the phase center of every individual sub-array. The sub-arrays boundaries are also outlined. Fig. \ref{fig:ims-fig3}b shows the phase shifters granularity for the 256-elements phased-array aperture with vertical and horizontal dotted line indicating that there are more indices along the x-axis (23 indices) than the y-axis (16 indices). Randomness is introduced within the array aperture as the locations of phase centers are not in a regular pattern.

\begin{figure}[!b]
\centering
\subfloat[]{
\label{f1}\includegraphics[width=0.24\textwidth]{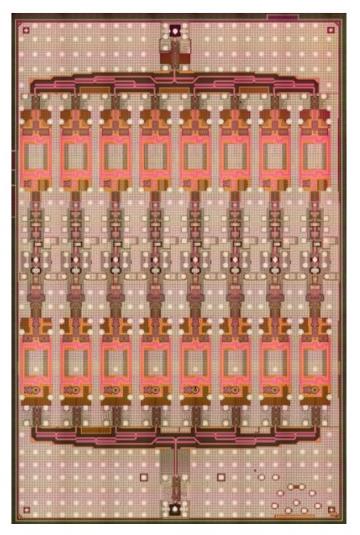}}
\subfloat[]{
\label{f2}\includegraphics[width=0.24\textwidth]{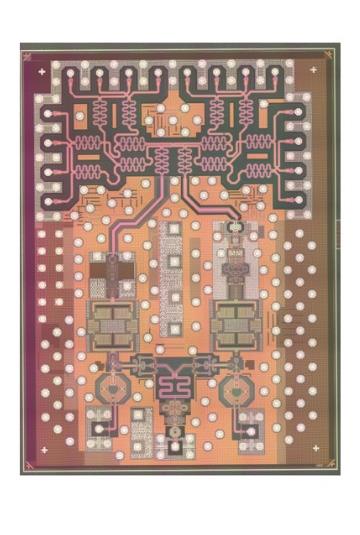}}
\caption{Photomicrograph of (a) Active Antenna and (b) Up/Down Converter ASICs}
\label{fig:ims-fig4}
\end{figure}


\begin{figure}[!b]
\centering
\subfloat[]{
\label{f1}\includegraphics[width=0.48\textwidth]{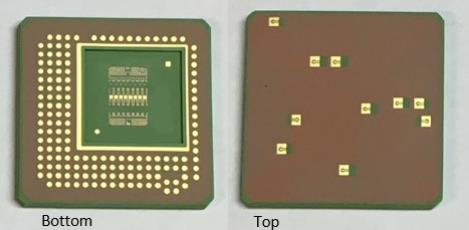}}\\
\subfloat[]{
\label{f2}\includegraphics[width=0.48\textwidth]{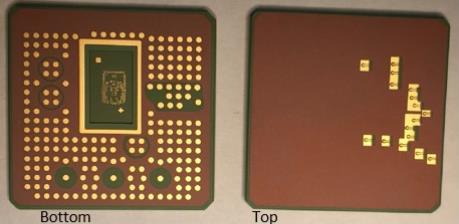}}
\caption{ Photomicrograph of (a) Active Antenna ($ 20 \times 20 \text{mm}^2 $) LTCC package and (b) Up/Down Converter ($ 20 \times 20 \text{mm}^2 $) LTCC package}
\label{fig:ims-fig5}
\end{figure}


 Photomicrograph of the two ASICs is shown in Fig. \ref{fig:ims-fig4}. They were both designed using the STMicroelectronics 55 nm SiGe BiCMOS process. LTCC packages were designed and manufactured to test the two ASICs separately as shown in Fig. \ref{fig:ims-fig5}. Only building block measurements were performed at the wafer level.  These LTCCs have the same stack-up as the antenna LTCC so all the transitions are identical.

\section{Calibration Method for Bore-Sight}\label{cali_1}
Due to non-ideal circuits and process variations, the varying electrical amplitude and phase of different sub-arrays need to be calibrated for beam-forming. For the bore-sight (elevation $\beta  = 0^o$ and azimuth $\alpha = 0^o$), the calibration method requires near field measurement and the signal processing. The near field measurement is done once to save time, which is around 40 minutes for our phased array antenna.


\subsection{Near Field Measurement}
\begin{figure*}[]
\centering
\subfloat[]{
\label{fig:measure_1}\includegraphics[width=0.57\textwidth]{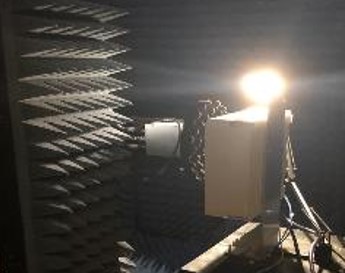}}
\ \ \ \
\subfloat[]{
\label{fig:measure_2}\includegraphics[width=0.35\textwidth]{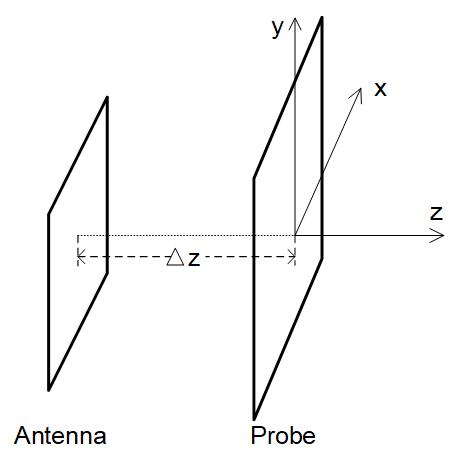}}
\caption{Near field measurement environment: (a) microwave chamber, and b) placement of phased array antenna and probe scanning in a rectangular plane}
\label{fig:chamber}
\end{figure*}

For the near field measurement, the phased array antenna is placed in a microwave chamber as shown in Fig.\ref{fig:measure_1}. The position of the phased array antenna and probe are shown in Fig.\ref{fig:measure_2}, where the probe moves in a rectangular plane with center at $(x,y,z) = (0,0,0)$, and the center of the antenna is $(\Delta x, \Delta y, \Delta z)$. The distance $\Delta z$ can be easily measured by Vernier calipers, since the antenna and probe are close to each other. The center shift $\Delta x$ and $\Delta y$ caused by placement error is not important. 

The antenna and the probe are set to have the same polarization, and the sampling spacing of the probe is less than  $\lambda/2$  \cite{sam_theo}.  During scanning as the probe moves, the amplitude and phase of each sub-array use default values with $0$ dB gain and $0^o$ phase. Then the scanning result is defined as $H_p(x,y,0)$, where $(x,y,0)$ is the probe position. We note that, the E-band signal received by the probe could be down-converted to low frequency and recorded as $H_p(x,y,0)$, to reduce the instrument performance requirement.

\subsection{NFFF and FFNF Transform}

Similar to the method in \cite{papers_1, papers_2}, with the scanning results, the NFFF transform is used to get the far field radiation performance $\tilde{R}(\alpha, \beta)$ of the phased sub-array antenna as:
\begin{align}
\tilde{R}(\alpha, \beta) = \frac{1}{2\pi}\int_{-\infty}^{\infty}\int_{-\infty}^{\infty}H_p(x,y,0)e^{j(k_x x + k_y y)}dxdy
\end{align}
where $k_x = 2\pi\sin(\alpha)\cos(\beta)/\lambda$ and $k_y = 2\pi\sin(\beta)/\lambda$.

Considering the non-ideal measurement with limited scanning range, the far field radiation performance of a reference antenna is achieved as $R_{ref}(\alpha, \beta)$. Let $R_{ref}^{real}(\alpha, \beta)$ be the real far field radiation performance of the reference antenna, the far field radiation performance $\tilde{R}(\alpha, \beta)$ of the phased sub-array antenna is revised as:
\begin{align}
{R}(\alpha, \beta) = \tilde{R}(\alpha, \beta) \cdot \frac{R_{ref}^{real}(\alpha, \beta)}{R_{ref}(\alpha, \beta)}
\end{align}

\begin{figure}[]
\centering
\subfloat[]{
\label{fig:amp_1}\includegraphics[width=0.25\textwidth]{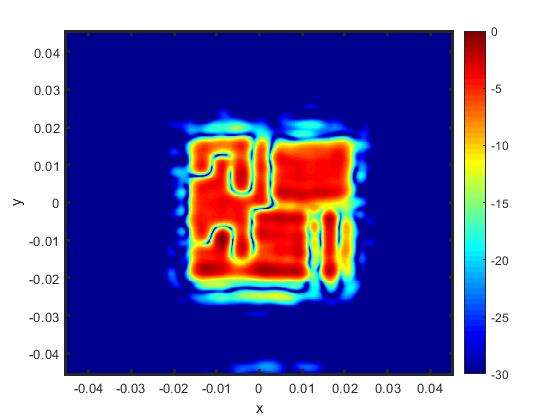}}
\subfloat[]{
\label{fig:phase_2}\includegraphics[width=0.25\textwidth]{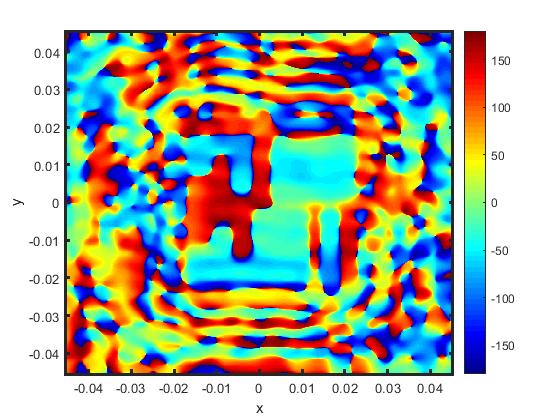}} \\
\subfloat[]{
\label{fig:amp_3}\includegraphics[width=0.25\textwidth]{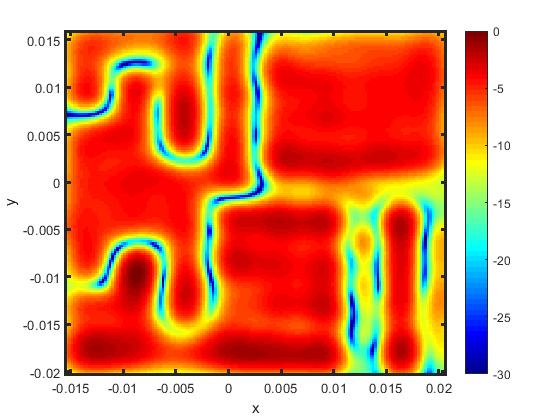}}
\subfloat[]{
\label{fig:phase_4}\includegraphics[width=0.25\textwidth]{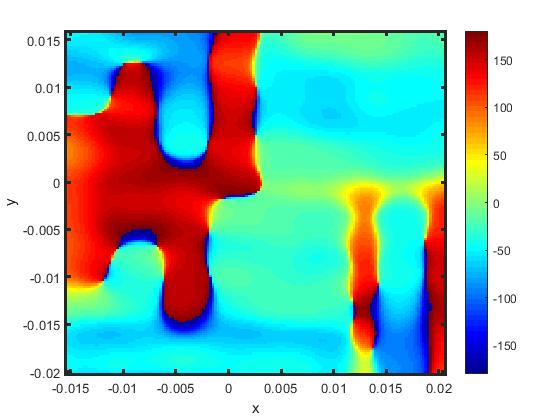}}
\caption{Radiation performance on the surface of 256-elements antenna before calibration: (a) amplitude, (b) phase, (c) extracted amplitude, and (d) extracted phase.}
\label{fig:rad_on_antenna}
\end{figure}

Then the FFNF transform is utilized to get the near field radiation performance on the antenna surface shown  with  amplitude in Fig. \ref{fig:amp_1} and phase in Fig. \ref{fig:phase_2} as:
\begin{align}
& H_{a}(x,y,-\Delta z)  \notag\\
& = \frac{1}{2\pi}\int_{-\infty}^{\infty}\int_{-\infty}^{\infty}{R}(\alpha, \beta) e^{j(-k_x x + -k_y y + k_z \Delta z)}d\theta d\phi
\end{align}
where $k_z = 2\pi\cos(\alpha)\cos(\beta)/\lambda$.

%

Let $S$ be the pre-known shape of the antenna, whose center is $(\hat{x},\hat{y})$. When $S$ moves, the amplitude in $S$ is summarized and the maximum one appears when $S$ overlaps the antenna.  Hence the center of the antenna $(x_0, y_0)$ is identified as :
\begin{align}
(x_0,y_0) = \max_{\hat{x},\hat{y}}\int\int_{(x,y)\in S_{(\hat{x},\hat{y})}} |H_{a}(x,y,-\Delta z)| dx dy
\end{align}
Then the radiation performance on the surface of 256-elements antenna is extracted as shown in Fig. \ref{fig:amp_3} and  Fig. \ref{fig:phase_4}. It matches roughly with the sub-array structure shown in  Fig. \ref{fig:ims-fig3}a. The sub-arrays are out of phase, which leads to the amplitude troughs between sub-arrays.


\subsection{Removing the Adjacent Influence}
The radiation performance of each sub-array is influenced by the adjacent sub-arrays. To further increase the calibration accuracy, this influence should be removed.
The phase calibration is described, and the amplitude calibration follows the same methodology.

Using the sub-array structure in  Fig. \ref{fig:ims-fig3}a, the position and area of each sub-array in  Fig. \ref{fig:phase_4} is identified. Then the phase of all sub-arrays are extracted. In general, let $M$ be the number of sampled points for each sub-array, where $M = 800$ is used for our antenna.

A two steps iterative procedure is used to remove the influence of adjacent sub-arrays. Firstly, the phase distance $d_{i,j}$ between any two points $0 < i, j < M$ with phase value $p_i$ and $p_j$ are calculated as:
\begin{align}
d_{i,j} = \min &\left\{  \mod (p_i - p_j + 360, 360), \right.\notag \\
 & \ \ \ \ \left. 360 - \mod (p_i - p_j + 360, 360)  \right\}
\end{align}
and the maximum distance is between points $\hat i$ and $\hat j$ as:
\begin{align}
d_{\hat i, \hat j} = \max_{0 < i, j < M} d_{i,j}.
\end{align}
Secondly, the one with larger distance between the remaining points will be removed, whose index is:
\begin{align}
\hat k = \max_{k \in \{\hat i, \hat j\}} \sum_{0<l<M, l \notin \{\hat i, \hat j\}} d_{k, l}
\end{align}
Then the procedure goes back to the first step until the maximum distance is smaller than a pre-defined threshold.

\begin{figure}[]
\centering
\includegraphics[width=0.5\textwidth]{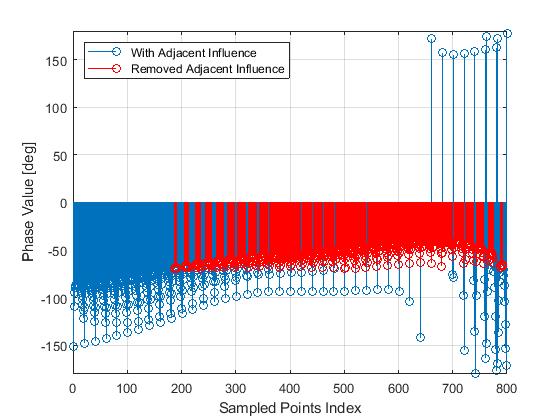}
\caption{Phase value of sampled points for one sub-array  before and after removing the adjacent influence.}
\label{fig:phase_all}
\end{figure}

With large threshold, more points will be remained with large variance. With small threshold, less points will be remained with small variance. After many experiments,  the threshold $30^o$  is utilized in our method. Take a sub-array as an example, whose phase value before and after removing the adjacent influence are shown in FIg. \ref{fig:phase_all}. It can be seen that,  after removing the bad points, the distance between any two of the remaining points is smaller than the threshold.

\begin{figure}[]
\centering
\includegraphics[width=0.5\textwidth]{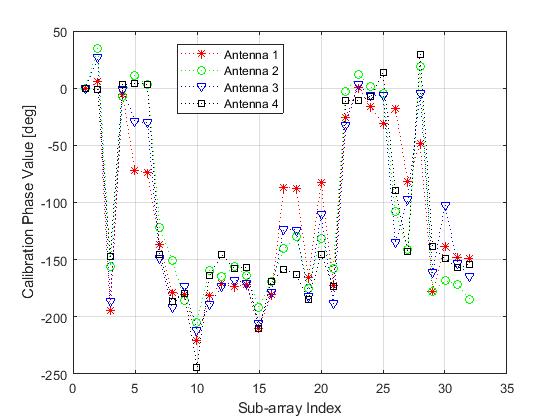}
\caption{The calibration phase value of the 32 sub-arrays of four 256-elements antennas.}
\label{fig:phase_cali}
\end{figure}




\begin{figure}[]
\centering
\includegraphics[width=0.5\textwidth]{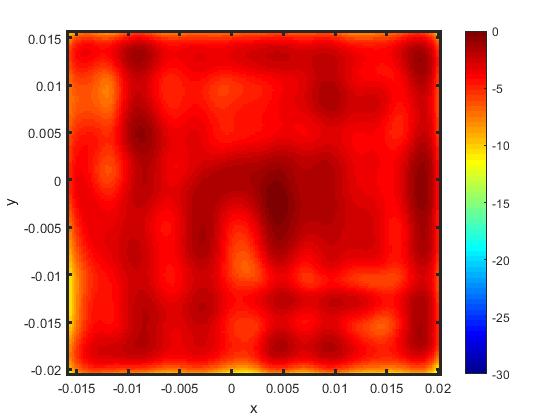}
\caption{Extracted  amplitude on the surface of 256-elements antenna  after calibration.}
\label{fig:rad_on_antenna2}
\end{figure}

\subsection{Calibration}

After removing the influence of adjacent sub-arrays, the average of the remaining points is regarded as the phase at bore-sight as:
\begin{equation}
p^k(0,0) = \sum_{1< i < M \ \& \text{  point i is left}} p_i / N  \label{equat_16}
\end{equation}
where $M$ is the number of remaining points. In general, $p^k(\alpha, \beta)$ is defined as the phase at direction $(\alpha, \beta)$ for the $k$-th sub-array. A simple calibration method is to maximize the power at bore-sight, and the calibration phase value for each sub-array is $- p^k(0,0)$.

The calibration phase value of the 32 sub-arrays of four 256-elements antennas are shown in Fig. \ref{fig:phase_cali}. It can be seen that, for the same antenna, the phase value of the 32 sub-arrays vary dramatically. For the same sub-array of different antennas, their phase value are also different due to the non-ideal circuits and process variations.

To verify the effectiveness, the calibration phase value of the 32 sub-arrays are loaded to the phase shifters. Following the same way, the radiation performance on  the surface of 256-elements antenna after calibration is shown in Fig.\ref{fig:rad_on_antenna2}. Compared with that before calibration in Fig.\ref{fig:amp_3}, the amplitude troughs between sub-arrays almost disappear, since the phase value of adjacent sub-arrays are almost the same.


\begin{figure}[]
\centering
\subfloat[]{
\label{fig:sum_1_8}\includegraphics[width=0.25\textwidth]{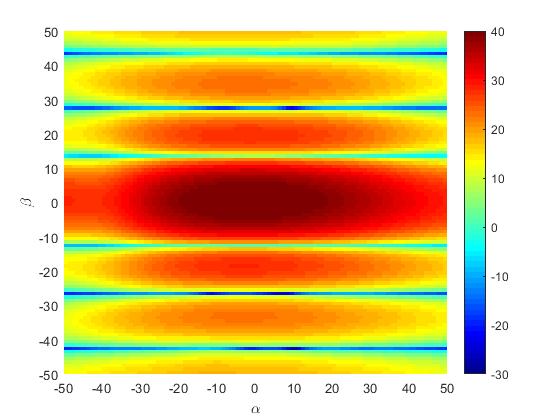}}
\subfloat[]{
\label{fig:phase_1_8}\includegraphics[width=0.25\textwidth]{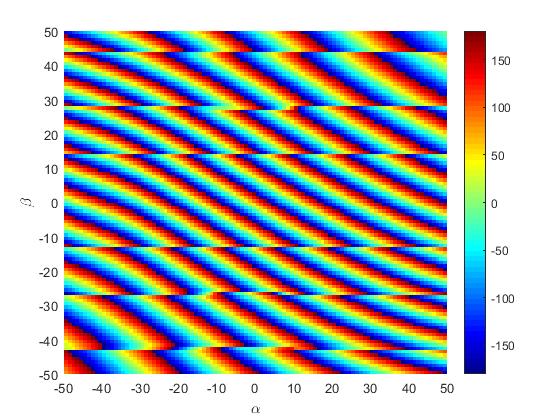}} \\
\subfloat[]{
\label{fig:amp_2_4}\includegraphics[width=0.25\textwidth]{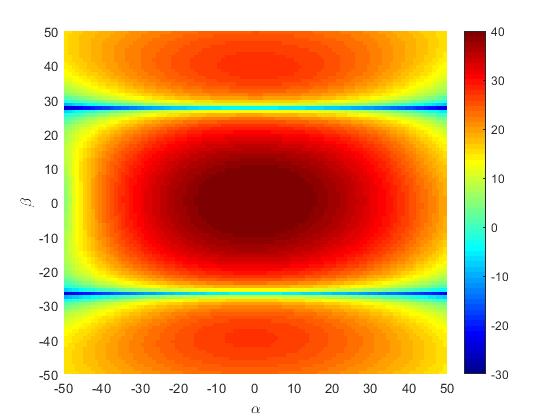}}
\subfloat[]{
\label{fig:phase_2_4}\includegraphics[width=0.25\textwidth]{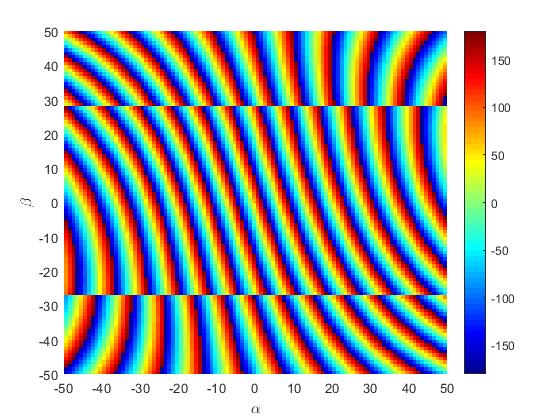}} \\
\subfloat[]{
\label{fig:amp_4_2}\includegraphics[width=0.25\textwidth]{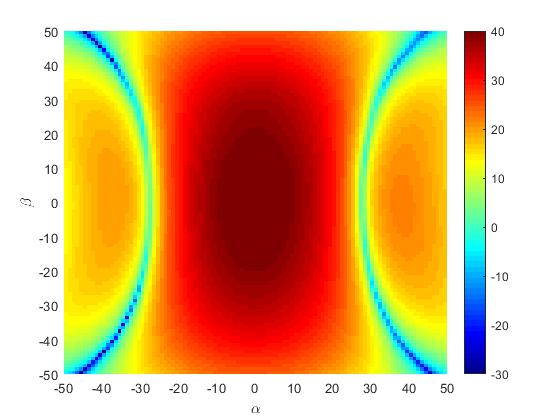}}
\subfloat[]{
\label{fig:phase_4_2}\includegraphics[width=0.25\textwidth]{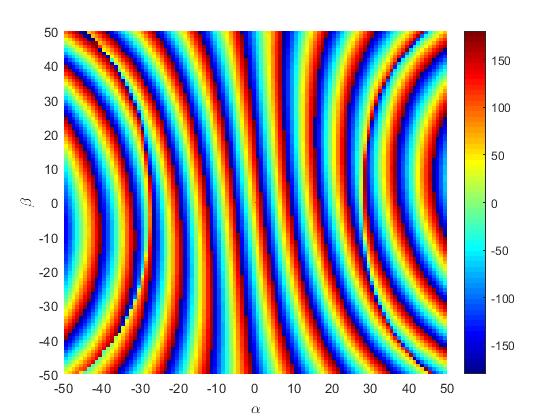}} \\
\subfloat[]{
\label{fig:amp_8_1}\includegraphics[width=0.25\textwidth]{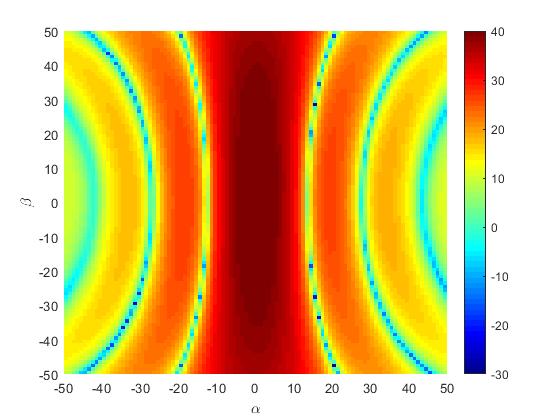}}
\subfloat[]{
\label{fig:phase_8_1}\includegraphics[width=0.25\textwidth]{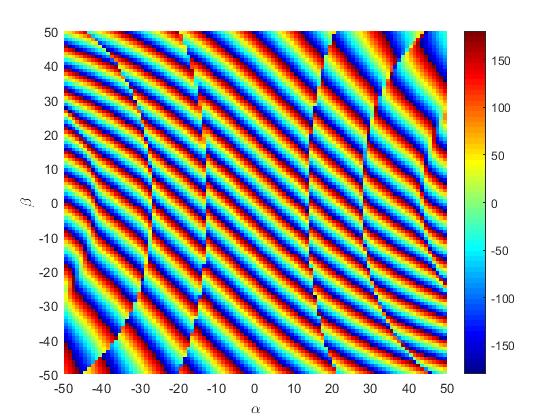}}
\caption{Far field radiation performance based on the HFSS simulations of the $4$ different shapes sub-arrays in our antenna: (a) amplitude of $1 \times 8$ sub-array , b) phase of $1 \times 8$ sub-array, (c) amplitude of $2 \times 4$ sub-array , (d) phase of $2 \times 4$ sub-array, (e) amplitude of $4 \times 2$ sub-array , (f) phase of $4 \times 2$ sub-array, (g) amplitude of $8 \times 1$ sub-array , and (h) phase of $8 \times 1$ sub-array.}
\label{fig:rad_on_hfss}
\end{figure}

\begin{figure}[]
\centering
\subfloat[]{
\label{fig:large_are}\includegraphics[width=0.5\textwidth]{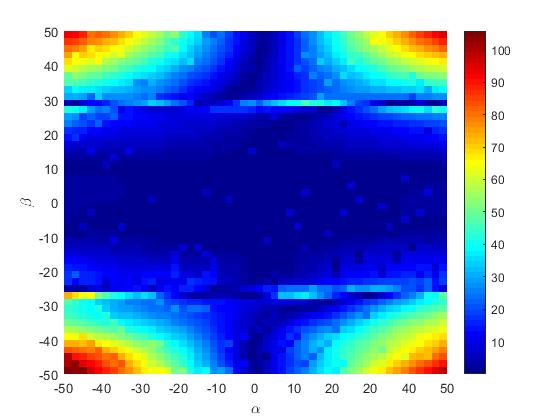}}\\
\subfloat[]{
\label{fig:interested_are}\includegraphics[width=0.5\textwidth]{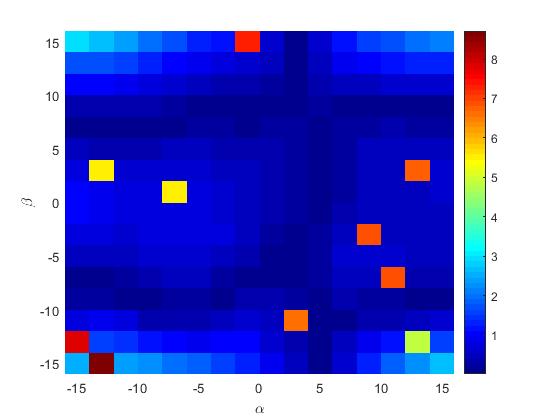}}
\caption{The phase difference of measured result and HFSS simulation result of one sub-array, minus the phase difference of measured result and HFSS simulation result of another sub-array in degree. (a) large radiation range, (b) beam steering range.}
\label{fig:rad_cmp}
\end{figure}

%

\section{Calibration Method for Non-Bore-Sight}\label{cali_2}
For the non-bore-sight, the calibration method utilizes the bore-sight calibration results and the HFSS simulation results of all sub-arrays.  Otherwise, the near field measurement for each sub-array will consume $40\times 32$ minutes for the 256-elements antenna, which is un-acceptable. Similar to the bore-sight calibration, the phase calibration for non-bore-sight is described, and the amplitude calibration uses the same methodology.

\subsection{Extraction at Non-Bore-Sight}
The far field radiation performance of all sub-arrays are simulated by HFSS. The 256-elements antenna has $32$ sub-arrays with $4$ kinds of shapes as $1 \times 8$, $2 \times 4$, $4 \times 2$ and $8 \times 1$. We randomly choose one sub-array for each shape, and show its far field amplitude and phase performance in Fig.\ref{fig:rad_on_hfss}.


The phase performance of different sub-array from measured results have different phase shift caused by the non-ideal hardware as shown in equation (\ref{equat_16}), and different phase performance pattern due to position shift and different kind of shape.  However, after removing the phase shift caused by the non-ideal hardware, the phase performance of each sub-array from measured results and that from HFSS simulation results should be the same as:
\begin{align}
&\left[ p^k(\alpha, \beta) - p^k(0, 0) \right] - \left[p_{hfss}^k(\alpha, \beta) - p_{hfss}^k(0, 0)\right] =  \notag \\
&\left[ p^l(\alpha, \beta) - p^l(0, 0) \right] - \left[p_{hfss}^l(\alpha, \beta) - p_{hfss}^l(0, 0)\right]  \label{equal_relation}
\end{align}
where $p^k(\alpha, \beta)$ and $p^l(\alpha, \beta)$ are the phase at direction $(\alpha, \beta)$ for the $k$-th and $l$-th sub-array from measured results, and
$p_{hfss}^k(\alpha, \beta)$ and $p_{hfss}^l(\alpha, \beta)$ are the phase at direction $(\alpha, \beta)$ for the $k$-th and $l$-th sub-array from HFSS simulation results. We note that the left and right side may not be zero, since there may exist slight error to align the coordination of measured results and that of HFSS simulation results.

%

To verify the relationship of equation (\ref{equal_relation}), we randomly select two sub-arrays. The near field measurement and NFFF transform are used to get their phase performance from measured results. The HFSS simulation results are also simulated. Then the left side of equation (\ref{equal_relation}) using one sub-array, minus the right side of equation (\ref{equal_relation}) using another sub-array, whose value should be zero as shown in Fig.\ref{fig:rad_cmp}.  In Fig. \ref{fig:large_are}, there exists non-negligible error with large $(\alpha, \beta)$. The reasons are 1) during the near field measurement, the limited scanning range lose some information, 2) the amplitude performance is quite low with large $(\alpha, \beta)$, which also degrades the accuracy. Fortunately, for the beam steering range as shown in Fig. \ref{fig:interested_are}, It can be seen that most of the values are smaller than $3$ degree and some bad points are also smaller than $8$ degree. Hence we believe that the phase performance of each sub-array from measured results and that from HFSS simulation results are the same, especially within the beam steering range.

%

\subsection{Calibration}
Using equation (\ref{equal_relation}), the phase at direction $(\alpha, \beta)$ for each sub-array is:
\begin{align}
p^k(\alpha, \beta)  =   p^k(0, 0)  - \left[p_{hfss}^k(\alpha, \beta) - p_{hfss}^k(0, 0)\right]  + C \label{nonbore_relation}
\end{align}
where $C$ is a constant shift due to non-aligned coordination of measured results and that of HFSS simulation results. A simple calibration method is to maximize the power at non-bore-sight direction, and the calibration phase value for each sub-array is $- p^k(\alpha, \beta)$.

We note that with the phase condition and far field radiation performance of each sub-array, more advanced calibration principle can be used to suppress the side-lode. One can use the proposed scheme and advanced calibration principle/algorithom to further improve their antenna beam-forming performance. Since our antenna has $256$ elements, its beam-forming performance already satisfies the requirement using the simple calibration method. Moreover, the amplitude calibration follows the same method and omitted here to save space.


\begin{figure}[]
\centering
\includegraphics[width=0.45\textwidth]{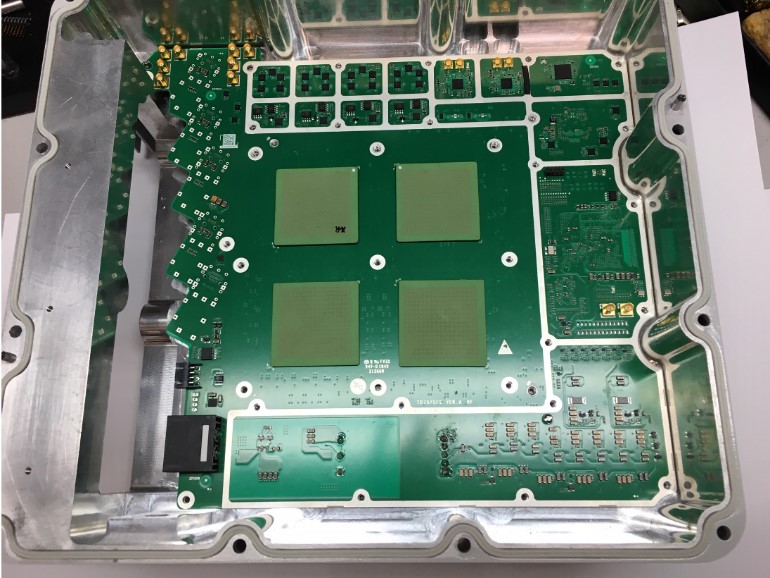}
\caption{Test system contains four 256-elements antennas.}
\label{fig:test_system}
\end{figure}

\section{Results}\label{results}


\subsection{Test System}

 The 256-elements antenna shown in Fig.\ref{fig:ims-fig2} has  32 sub-arrays on the top side, and four AA ASIC and one UDC ASIC on the bottom side. The AA ASIC includes eight TRX channels, mm-wave splitter/combiner and associated control for the LNA, PA, 6-bit PS \cite{ims2, ims3, ims4} and variable gain amplifiers controlled using an SPI interface. The UDC ASIC receives/transmits IF signal from the baseband module, and upconverts/downconverts E-Band to/from AA. Each UDC feeds the four AAs, and each AA feeds eight sub-arrays.

 The test system contains four 256-elements antennas as shown in Fig.\ref{fig:test_system}. The field programmable gate array (FPGA) sequential logic is developed to control the antenna through SPI interface. The phase and gain registers of each sub-array are used for beam steering, and the switch register is utilized for TRX switch. The SPI clock speed is larger 100 MHz for fast configuring the registers. After configuration, the time required for the beam to be stable is less than 5 ns, which is much smaller than the cyclic prefix length of NR numerology \cite{5g-nr-cp}. Hence, the antenna supports the beam switch per orthogonal frequency-division multiplexing (OFDM) symbol sequentially in the real system.

 \subsection{Beam-Forming Performance}
 For the beam-forming performance, the test system is placed in the chamber as shown in Fig.\ref{fig:measure_1}.

 \begin{figure}[]
\centering
\subfloat[]{
\label{fig:far_before_1}\includegraphics[width=0.26\textwidth]{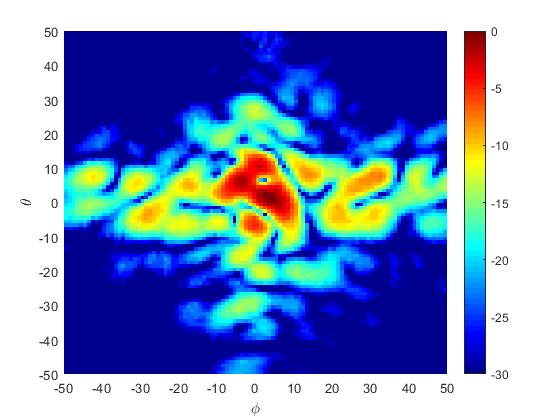}}
\subfloat[]{
\label{fig:far_before_2}\includegraphics[width=0.26\textwidth]{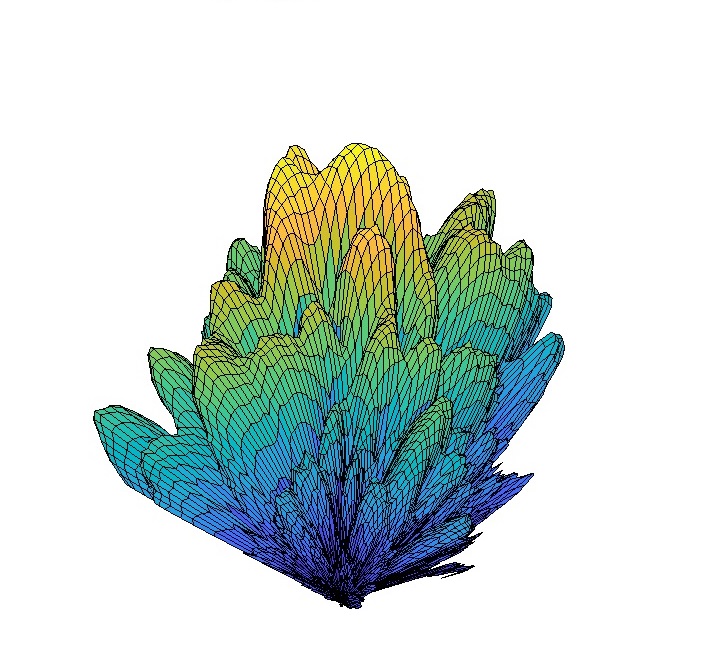}}
\caption{Far-field radiation performance of the 256-elements antenna at 72.2GHz before calibration: (a) normalized amplitude, (b) 3D radiation shape.}
\label{fig:far_before_antenna}
\end{figure}

 \subsubsection{Without Calibration}
 Before calibration, default values with $0$ dB gain and $0^o$ phase are loaded to the registers. The probe scans once to get the near field radiation performance on the probe panel. Then using NFFF transform, the far field radiation  performance is obtained as shown in Fig. \ref{fig:far_before_antenna}. It can be seen that, due to the non-ideal circuit and inconsistence of RF channels, the beam is non-shaped. Then the calibration method is needed for beam forming to increase the antenna gain.

 \subsubsection{Beam-forming Performance of Bore-Sight After Calibration}
 Using the previous near field measurement, NFFF transform and FFFF transform, the radiation performance on the antenna surface is obtained. Based on the sub-array shape, the phase of all sub-arrays are extracted and the adjacent influence is removed. Then the calibration phase value for bore-sight is $0^o$ minus the average phase for each sub-array.

\begin{figure}[]
\centering
\subfloat[]{
\label{fig:far_after_bore_1}\includegraphics[width=0.26\textwidth]{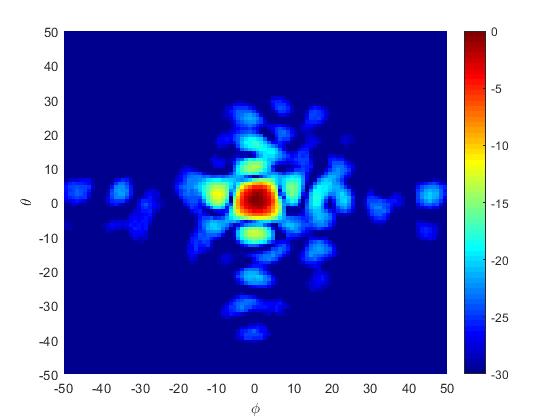}}
\subfloat[]{
\label{fig:far_after_bore_2}\includegraphics[width=0.26\textwidth]{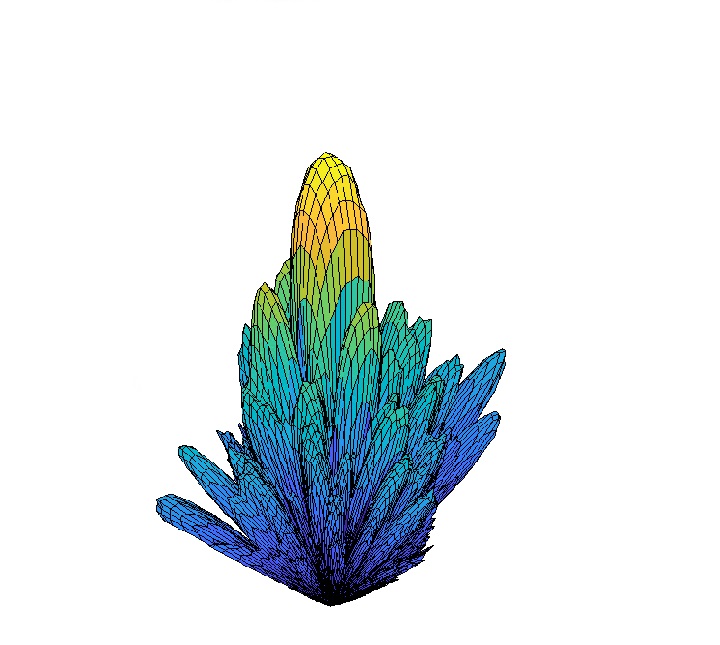}} \\
\subfloat[]{
\label{fig:far_after_bore_3}\includegraphics[width=0.25\textwidth]{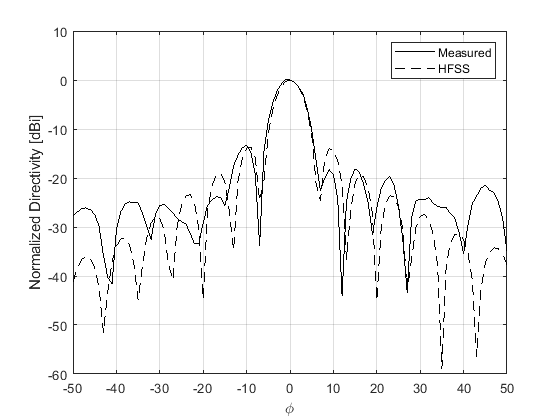}}
\subfloat[]{
\label{fig:far_after_bore_4}\includegraphics[width=0.25\textwidth]{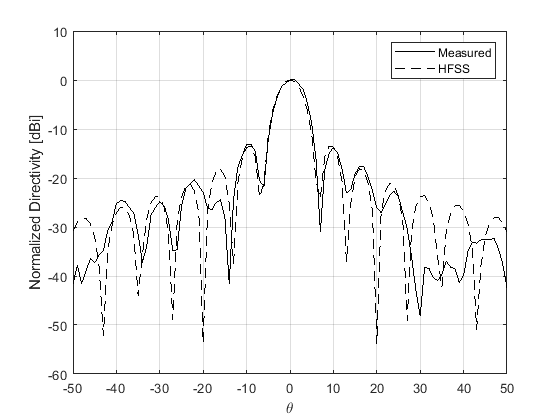}}
\caption{Far-field radiation performance of the 256-elements antenna at 72.2GHz and the main beam steers to the bore-sight after calibration: (a) normalized amplitude, (b) 3D radiation shape, (c) cut figure with elevation $\theta = 0^o$ , and (d) cut figure with azimuth $\phi = 0^o$.}
\label{fig:far_after_bore}
\end{figure}

 The calibration phase value of all sub-arrays are loaded to the registers. For our antenna with large number of elements, the performance with and without amplitude calibration are almost the same. Following the same method, after one time near field measurement and NFFF transform, the far field radiation performance is shown in Fig. \ref{fig:far_after_bore}. It can be seen that, for the bore-sight beam steering,  the measured result after calibration is shaped and in good agreement with the HFSS simulation result, where the SSL is about $13$ dBc below the main beam.
 
 \begin{figure}[]
\centering
\subfloat[]{
\label{fig:far_after_nonbore_1}\includegraphics[width=0.26\textwidth]{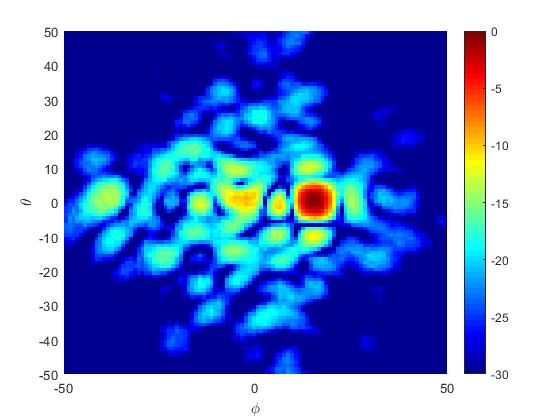}}
\subfloat[]{
\label{fig:far_after_nonbore_2}\includegraphics[width=0.26\textwidth]{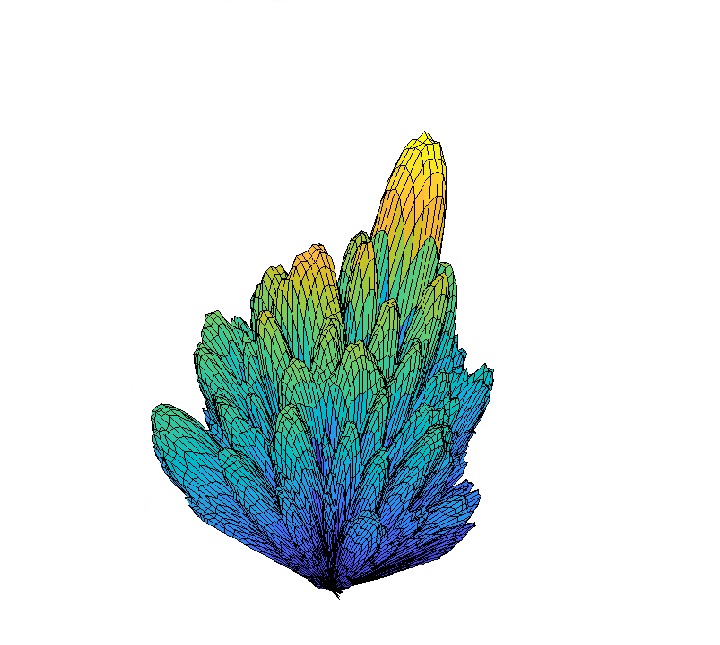}} \\
\subfloat[]{
\label{fig:far_after_nonbore_3}\includegraphics[width=0.25\textwidth]{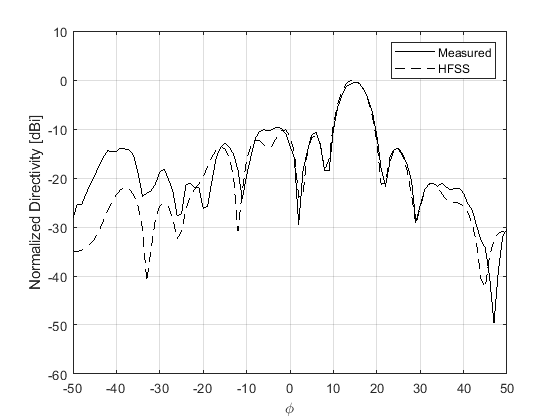}}
\subfloat[]{
\label{fig:far_after_nonbore_4}\includegraphics[width=0.25\textwidth]{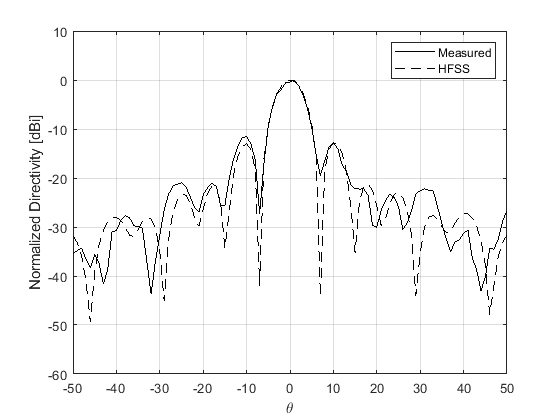}}
\caption{Far-field  radiation performance of the 256-elements antenna at 72.2GHz and the main beam steers to non-bore-sight direction $(0^o,15^o)$ after calibration: (a) normalized amplitude, (b) 3D radiation shape, (c) cut figure with elevation $\theta = 0^o$ , and (d) cut figure with azimuth $\phi = 15^o$.}
\label{fig:far_after_nonbore}
\end{figure}

  \subsubsection{Beam-forming Performance of non-bore-sight After Calibration}
 After removing the phase shift caused by the non-ideal hardware, the phase performance of each sub-array from measured results and that from HFSS simulation results are the same. Then using  the calibration phase value of bore-sight and the HFSS simulation results, the calibration phase value for non-bore-sight is obtained by equation (\ref{nonbore_relation}) for each sub-array.


 The calibration phase value of all sub-arrays are loaded to the registers.  Following the same method, after one time near field measurement and NFFF transform, the far field radiation performance at non-bore-sight direction $(0^o,15^o)$ is shown in Fig. \ref{fig:far_after_bore}. It can be seen that, for the bore-sight beam steering,  the measured result after calibration is shaped and in good agreement with the HFSS simulation result, where the SSL is about $10$ dBc below the main beam.


 With arbitrary sub-array structure, a straightforward calibration method is to do the near field measurement for each sub-array individually. Then NFFF transform is used to get the far field radiation performances of different sub-arrays. For a specific steering direction, their difference can be utilized for the calibration. However, the number of near field measurement equals the number of sub-arrays, which is time consuming.  The proposed calibration method requires only a single near-field measurement, which is much faster and saves time.

\begin{table}[]
\centering
\caption{Summary of measurement results over 1 meter with one 256-elements antenna at one end and a standard horn antenna at the other.}
\label{tb1}
\renewcommand{\arraystretch}{1.0}
\begin{tabular}{p{0.19\textwidth}|p{0.12\textwidth}|p{0.1\textwidth}}
\hline
&TX Mode & RX Mode\\
\hline
Conversion gain &10.5 dB & 30 dB \\
\hline
P1dB (per PA) & 9.5 dBm &  \\
\hline
OIP3 (per PA) & 18.2 dBm &  \\
\hline
ACP(@Pin = -20 dBm) &-31.2/-30.8 dB & \\
\end{tabular}
\begin{tabular}{p{0.19\textwidth}|p{0.243\textwidth}}
\hline
LO Power @ 20.55GHz & 3 dBm \\
\hline
Matching (RF,IF,LO) & $< -20$ dB \\
\hline
\end{tabular}
\begin{tabular}{p{0.19\textwidth}|p{0.12\textwidth}|p{0.1\textwidth}}
Power consumption  (4AA + 1UDC) &4.2 W & 2.2 W \\
\hline
\end{tabular}

\end{table}

\begin{figure}[]
\centering
\includegraphics[width=0.5\textwidth]{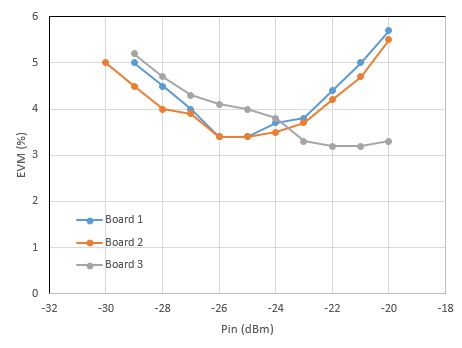}
\caption{The EVM versus the input power for a 1.5 GHz bandwidth 64-QAM signal using one 256-elements antenna as Tx and a standard horn antenna as Rx.}
\label{fig:evm}
\end{figure}

\subsection{Over the Air (OTA) Measurement}

Over a distance of one meter, the OTA measurement is done with one 256-elements antenna at one end and a standard horn antenna at the other. A summary of the test results of the 256-elements antenna as transmitter or receiver are shown in Table \ref{tb1}. The EVM is also measured with one 256-elements antenna as transmitter and the standard horn antenna as receiver, where the EVM versus the input power for a 1.5 GHz bandwidth 64-QAM signal is shown in Fig. \ref{fig:evm}. Due to non-ideal hardware, the EVM varies for different 256-elements antennas. Hence the input power of each 256-elements antenna should be carefully optimized to minimize the EVM.

\begin{figure}[]
\centering
\includegraphics[width=0.5\textwidth]{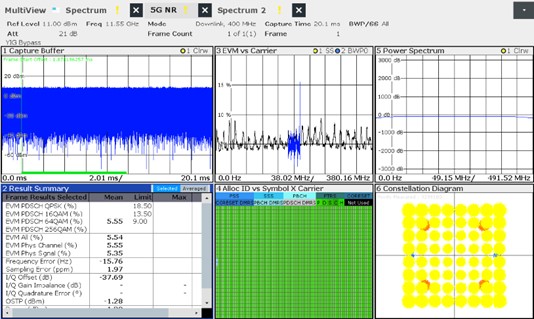}
\caption{Measured 64-QAM constellation at 20m link distance.}
\label{fig:64qam}
\end{figure}

In the office environment over two different distances 10 meter and 20 meter, the OTA measurement is done with two 256-elements antennas as transmitter and receiver. A summary of the EVM with different modulation schemes are shown in Table \ref{tb2}.  A screen capture showing the measured constellation at 20 meter for the 64-QAM signal is shown in Fig. \ref{fig:64qam}.
During the measurement,  a 5G NR, 400 MHz bandwidth, OFDM waveform with 120 kHz sub-carrier spacing is used. The lack of equipment prevents higher bandwidth signal to be used, but the office environment measurement suggests that more than 400 MHz bandwidth signal is possible.


\begin{table}
\centering
\caption{EVM OTA measurements between two 256-elements antennas in the office environment.}
\label{tb2}
\renewcommand{\arraystretch}{1.0}
\begin{tabular}{p{0.045\textwidth}|p{0.045\textwidth}|p{0.045\textwidth}|p{0.045\textwidth}p{0.045\textwidth}p{0.045\textwidth}p{0.045\textwidth}}
\hline
EIRP &Length & \small{RX Power} & EVM &  &  &  \\
\end{tabular}
\begin{tabular}{p{0.045\textwidth}|p{0.045\textwidth}|p{0.045\textwidth}|p{0.045\textwidth}|p{0.045\textwidth}|p{0.045\textwidth}|p{0.045\textwidth}}
\hline
 &  & &  64-QAM & QPSK & 16-QAM & 256-QAM  \\
\hline
dBm & m & dBm & (\%) & (\%) & (\%) & (\%) \\
\hline
33.5 & 10 & -28.3 & 4.31 & 4.38 & 4.28 & 4.84 \\
\hline
35.5 & 20 & -34.2 & 5.55 & 5.58 & 5.66 & 6.13 \\
\hline
\end{tabular}
\end{table}

\section{Conclusion}\label{conclusion}

This paper presents a highly-integrated E-band phased-array transceiver front-end in SiGe BiCMOS and LTCC, which is designed, manufactured, calibrated, measured in the chamber and the office environment.  A near-field measurement, NFFF transform, FFNF transform and comparison with HFSS simulation results is utilized for bore-sight and non-bore-sight calibration. The calibration method reduces the calibration time from $40 \times 32$ minutes to $40$ minutes. Measured in the chamber, the beam forming performance after calibration is highly improved and in good agreement with the simulations. Measured in the office environment, the 256-elements antenna achieves good EVM performance with 400MHz bandwidth, where the bandwidth can be further increased.  The proposed 256-elements antenna design, manufacture and calibration methods can be easily generalized to the antenna with arbitrary frequency and sub-array structure.

\section*{Conflict of Interests}
The authors declare that there is no conflict of interests regarding the publication of this paper.


\end{document}